\documentclass[sigconf]{aamas}

\usepackage{balance} 
\usepackage{xcolor}
\usepackage{algorithm}
\usepackage{algpseudocode}
\usepackage{amsmath}
\usepackage{float}
\usepackage{multirow}
\usepackage[table]{xcolor} 
\usepackage{siunitx}

\setcopyright{ifaamas}
\acmConference[AAMAS '26]{
}{
}
{
}{
}
\copyrightyear{
}
\acmYear{
}

\title[AAMAS-2026 Formatting Instructions]
{Autonomous vehicles need social awareness to find optima in multi-agent reinforcement learning routing games.}

\author{Anastasia Psarou}
\affiliation{
  \institution{Jagiellonian University, Faculty of Mathematics and Informatics}
  \city{Krakow}
  \country{Poland}}
\email{anastasia.psarou@doctoral.uj.edu.pl}

\author{Łukasz Gorczyca}
\affiliation{
  \institution{Jagiellonian University, Faculty of Mathematics and Informatics}
  \city{Krakow}
  \country{Poland}}
\email{lukasz.gorczyca@student.uj.edu.pl}

\author{Dominik Gaweł}
\affiliation{
  \institution{University of Warsaw, Faculty of Mathematics, Informatics, and Mechanics}
  \city{Warsaw}
  \country{Poland}}
\email{dg448617@students.mimuw.edu.pl}

\author{Rafał Kucharski}
\affiliation{
  \institution{Jagiellonian University, Faculty of Mathematics and Informatics}
  \city{Krakow}
  \country{Poland}}
\email{rafal.kucharski@uj.edu.pl}

\begin{abstract}
Previous work has shown that when multiple selfish Autonomous Vehicles (AVs) are introduced to future cities and start learning optimal routing strategies using Multi-Agent Reinforcement Learning (MARL), they may destabilize traffic systems, as they would require a significant amount of time to converge to the optimal solution, 
equivalent to years of real-world commuting.

We demonstrate that moving beyond the selfish component in the reward significantly relieves this issue. If each AV, apart from minimizing its own travel time, aims to reduce its impact on the system, this will be beneficial not only for the system-wide performance but also for each individual player in this routing game.

By introducing an intrinsic reward signal based on the marginal cost matrix, we significantly reduce training time and achieve convergence more reliably. 
Marginal cost quantifies the impact of each individual action (route-choice) on the system (total travel time). Including it as one of the components of the reward can reduce the degree of non-stationarity by aligning agents' objectives. Notably, the proposed counterfactual formulation preserves the system's equilibria and avoids oscillations.

Our experiments show that training MARL algorithms with our novel reward formulation enables the agents to converge to the optimal solution, whereas the baseline algorithms fail to do so. We show these effects in both a toy network and the real-world network of Saint-Arnoult. Our results optimistically indicate that social awareness (i.e., including marginal costs in routing decisions) improves both the system-wide and individual performance of future urban systems with AVs.

\end{abstract}


\newcommand{\BibTeX}{\rm B\kern-.05em{\sc i\kern-.025em b}\kern-.08em\TeX}

\begin{document}


\pagestyle{fancy}
\fancyhead{}


\maketitle 


\section{Introduction}
Every day, selfish human drivers \cite{WARDROP1952} make routing decisions to travel from an origin to a destination through a traffic network. For instance, they commute from home to work by selecting among multiple possible routes, typically the one with the lowest expected travel time \cite{inductive_bias_bounded_rationality}. 
The anticipated integration of Autonomous Vehicles (AVs) into future transportation systems will introduce new dynamics, as AVs may not only make independent routing decisions but also possess capabilities for coordination, learning, or communication \cite{FAGNANT2015167}. Unlike human drivers, the objectives of AVs may not necessarily be selfish. To model AV routing decisions, Multi-Agent Reinforcement Learning (MARL) has emerged as a promising method, with previous research already exploring this application \cite{AgentRewardShaping, lazar2021learningdynamicallyrouteautonomous, Thomasini+2023, akman2024impact}. Yet despite its promise, MARL presents challenges when applied to real-world traffic environments. Prior research has shown that when multiple \textit{selfish} AVs simultaneously learn routing strategies using state-of-the-art MARL algorithms, they may destabilize traffic networks \cite{psarou2025collaborationcitymachinelearning}. This can happen because the algorithms used in that work require long training iterations, equivalent to several years of real-world commuting, to converge to the optimal solution (system optimal or individually optimal), or in some cases, fail to converge at all. 

\begin{figure*}[!t]
\begin{center}
\centerline{\includegraphics[width=1\textwidth]{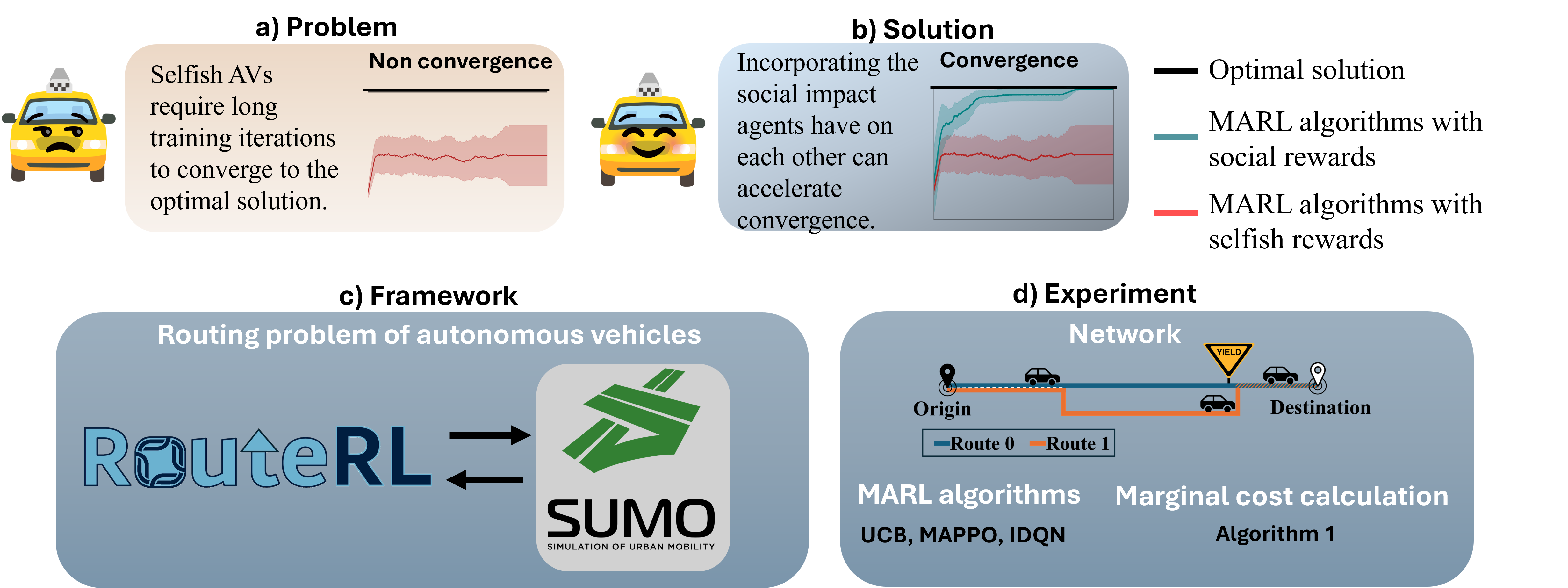}}
\caption{Overview: When multiple selfish AVs are introduced into future cities and have to simultaneously learn optimal routing strategies using MARL they need long training iterations to converge to the optimal solution (a). However, if we incorporate the impact an AV's presence has on the other agents of the system, into its travel-time based reward value, can improve the convergence of MARL algorithms applied to the route choice problem of AVs (b). This enables faster convergence to the system-optimal and is can also be more beneficial for individual AV agents. We show this with experiments using RouteRL \cite{akman2025routerlmultiagentreinforcementlearning}, a Multi-Agent Reinforcement Learning (MARL) framework that models the routing decisions of AVs and human drivers (c), and demonstrate our results using the \textbf{T}wo-\textbf{R}oute (\textbf{Y}ield) network (\textit{TRY}), d), where agents choose between \textit{Route 0}, which is the shortest route without priority, and \textit{Route 1}, a slightly longer alternative with priority.}
\label{fig:overview}
\end{center}
\end{figure*}

To address this instability and enable faster convergence to the optimal solution, we propose an intrinsic reward signal that complements the standard \textit{selfish} travel-time-based reward. In intrinsically motivated Reinforcement Learning (RL) \cite{NIPS2004_4be5a36c}, the reward signal encourages agents to explore or learn behaviors beyond what is explicitly defined by selfish (external) rewards, like social behaviors. Jaques et al. \cite{jaques2019socialinfluenceintrinsicmotivation} demonstrated that rewarding agents for having causal influence over other agents' actions, using counterfactual reasoning, can promote coordination and communication between them. 
\textbf{In this work, we demonstrate that incorporating the impact an AV agent’s presence has on the other agents of the system into its reward value enables the system to reach the optimum solution faster.} Specifically, we show that:
\begin{itemize}
    \item \textbf{Incorporating a social component in the reward of the AV agents accelerates convergence to the optimal solution.} Specifically, including the marginal cost in the immediate reward of the AVs brings the system faster to the System Optimal (SO) solution (as demonstrated in Figs.  \ref{fig:overview}b, \ref{fig:try_network_results}), both in deterministic and non-deterministic traffic settings. This leads to more efficient solutions, as both AVs and human agents experience shorter average travel times, once the training process is complete, as shown in Table \ref{tab:avg_travel_times}.

    \item \textbf{The equilibria of the system remain unchanged as the coefficient of the counterfactual intrinsic reward increases.} This result demonstrates that, as the coefficient of the intrinsic reward increases,  convergence is accelerated towards the unchanged system optimal equilibria, as depicted in Figs. \ref{fig:reward_shaping}, \ref{fig:equilibria}.
    \item \textbf{Our results are demonstrated in a small and in a bigger real-world network.}  Initially, we show that AVs converge faster in the \textbf{T}wo-\textbf{R}oute (\textbf{Y}ield) network (\textbf{TRY}), depicted in Fig. \ref{fig:overview}d that is deliberately designed for studying this problem. In section \ref{sec:real_world_network}, we extend our method in the real-world Saint-Arnoult traffic network and show that the system and individual travel times are reduced when social AVs are introduced in the system (Table \ref{tab:av_travel_times_saint_arnoult}). 
    \item \textbf{Our method, specifically the way we compute the marginal cost matrix and incorporate it in the reward signal of an agent, is general and can be incorporated in any MARL algorithm for the routing of AVs.} 
    
\end{itemize}

\subsection{Related work}
\label{sec:related_works}

In recent years, AVs have attracted significant research interest \cite{li2024generalization, STERN2018205, yang2025public, ZHANG2024103887}. Studies suggest that their integration may improve road safety and reduce traffic congestion \cite{AbdelAty2024, Zheng2025}. Other studies highlight potential negative effects, such as increased emissions \cite{Onat2023}. However, limited research has examined the route choice decisions of AVs and their potential effects, both positive and negative, in urban environments.

Jamr{\'o}z et al. \cite{jamroz2025social} investigated the route choice problem for Connected Autonomous Vehicles (CAVs) and Human Driven Vehicles (HDVs) and demonstrated that CAVs, by optimizing different goals, can either benefit or systematically disadvantage human drivers. In their analysis, CAVs differ from HDVs in their capabilities to make collective decisions. Several works have explored the AV route choice problem with the use of MARL \cite{akman2024impact, akman2025urburbanrouting, AgentRewardShaping, lazar2021learningdynamicallyrouteautonomous, Thomasini+2023}. For instance, Akman et al. \cite{akman2025routerlmultiagentreinforcementlearning} developed RouteRL, a MARL framework that simulates the routing decisions of AVs in different traffic networks shared with human drivers. RouteRL uses SUMO \cite{SUMO2018}, a continuous, microscopic traffic simulator. Building on this work, Psarou et al. \cite{psarou2025collaborationcitymachinelearning} demonstrated that when multiple selfish AVs simultaneously learn optimal routing strategies using MARL, they may destabilize traffic networks by increasing travel times for both AVs and human drivers. 

MARL encompasses several learning paradigms. Independent learning (IL) is one of them, where each agent learns its own policy while treating the other agents as part of the environment \cite{iql}. A widely used RL algorithm is Deep Q-learning \cite{mnih2013playingatarideepreinforcement}, whose IL MARL variant is the Independent Deep Q-learning (IDQN). Another MARL learning paradigm is the Centralized Training Decentralized Execution (CTDE), where agents learn decentralized policies in a centralized way that are used independently in the execution phase. The Multi-Agent Proximal Policy Optimization (MAPPO) algorithm \cite{mappo} is a representative CTDE approach and is regarded as a baseline for cooperative MARL tasks \cite{papoudakis2021benchmarkingmultiagentdeepreinforcement}. Finally, in certain scenarios, bandit settings, each agent selects a single action per episode. A classic multi-armed bandit algorithm is the Upper Confidence Bound (UCB) algorithm \cite{auer2002finite}.

In congested road networks, each additional traveler imposes an external cost on others by increasing their travel times \cite{congestion_theory_transport_investment}. Peeta et al. \cite{peeta1995system} defined the Marginal Travel Time (MTT) of a time-dependent path as the impact that adding one vehicle to a given path at a specific timestep has on the overall system travel time in a macroscopic setting. Sheffi et al. \cite{sheffi1985urban} described the MTT of a link as the "marginal contribution of an additional traveler on the link to the total travel time on this link." 

Marginal cost pricing (MCT) has been studied as a means of improving traffic efficiency. The core idea is to charge each traveler a toll equal to the additional delay other drivers in the system experience because of the traveler's presence in the system \cite{COLE2006444}. 
Building on this concept, Ramos et al. \cite{toll_based_learning_minimizing_congestion} proposed a reinforcement learning algorithm that realigns heterogeneous preferences of drivers over travel time with the marginal cost imposed on the user as a toll. This tolling algorithm leads to a system-efficient equilibrium. Additionally, Sharon et al. \cite{marginal_cost_pricing_fixed_error_factor} studied the impact of the MCT and demonstrated that underestimating its value can result in system performance comparable to scenarios without any tolling. 

In this work, we incorporate the marginal travel time into the agent’s reward function; similar reward-shaping strategies have been adopted in prior studies addressing the routing problem of vehicles. For example, Grunitzki et al. \cite{grunitzki2014individual} studied a setting where drivers learn their route choices, using difference rewards. In the case of difference rewards, a default action is chosen, and the agent's reward is calculated as the difference between the travel time of the system and the travel time of the system when the agent chooses the default action. In their method, the difference reward of an agent is defined as the global system reward minus the global system reward when the specific agent is eliminated from the system. This approach is inspired by Tumer et al. \cite{AgentRewardShaping} that demonstrated that using difference utility rewards can lead agents to achieve outcomes that are not only beneficial for themselves but also more advantageous for the system as a whole. In both of these approaches, the agent's reward is computed by approximating the system's travel time and excluding the states where the specific agent has no impact.



Our approach is related to counterfactual techniques. In COMA-style counterfactual baselines, the credit term marginalizes over the agent’s actions while holding others fixed, which can be costly in large action spaces \cite{foerster2024counterfactualmultiagentpolicygradients}. In contrast, our method requires only a single counterfactual evaluation (presence versus absence), which makes it substantially more tractable in large action spaces. Additionally, our counterfactual preserves the set of equilibria in the system and improves stability \cite{reward_shaping_episodic_rl, theoretical_considerations_potential_based}, as we show in section \ref{sec:equilibria}.



To the best of our knowledge, no prior work has addressed the problem identified in \cite{psarou2025collaborationcitymachinelearning}: selfish AV agents, when learning their route choices using MARL, converge to suboptimal solutions that destabilize traffic networks. In particular, it remains unclear whether introducing social behavior to the AVs could mitigate this problem. By addressing this problem, we not only resolve the convergence issue but also enhance the societal impact, making systems with limited capacity more efficient and effective.

\section{Method}
We begin by introducing the simulation environment in section \ref{sec:simulation} within which, in section \ref{sec:problem_formulation}, we present the problem formulation. The immediate reward function of each agent is shown in section \ref{sec:intrinsic_reward_signal}, followed with formulation of the marginal travel time (MTT) and explanation how it is incorporated into the intrinsic reward structure of each AV agent in section \ref{sec:mtt}.

\subsection{Urban routing problem}
\label{sec:simulation}
To simulate the day-to-day routing decisions of human drivers and AVs, we utilize RouteRL \cite{akman2025routerlmultiagentreinforcementlearning}, a MARL framework integrated with SUMO \cite{SUMO2018}, a continuous microscopic traffic simulator that enables modeling complex urban traffic networks and agent interactions. Within RouteRL, drivers everyday (iteration) choose a route to travel from their origin to their destination; for example, traveling from home to work at a specific time. Feasible routes are computed by Janux \cite{JanuX}. Set of routes do not change during the simulation or from one day to the next, and both departure times and demand remain constant across days of simulation.

Within RouteRL, human drivers are represented as self-interested utility maximizers who iteratively adapt their route choices based on their experienced travel times \cite{Gawron1998, Cascetta2009}. Their learning process is assumed to guide the system toward a Wardrop equilibrium, a state in which no individual can improve their outcome by unilaterally changing routes \cite{WARDROP1952}. Once this equilibrium is reached, AVs are introduced into the network and use MARL algorithms to learn optimal routing policies through repeated interactions with the environment. 

Both AVs and human drivers interact with the environment by selecting a route from a predefined set of discrete alternatives to reach their destination. Agents’ decisions, due to limited capacity, contribute to system travel costs (congestion). This setting appears non-stationary from the perspective of a single agent, as the actions of the agents jointly influence the environment. AVs receive as observation the current state of the network, that is, the route choices of the vehicles that have already departed. Additionally, selfish AV agents receive a reward equal to the negative of their travel time upon arrival, and their objective is to maximize this reward value (see details in \cite{akman2025routerlmultiagentreinforcementlearning}). 

Using SUMO, we can simulate deterministic and non-deterministic traffic dynamics. In deterministic scenarios, traffic dynamics remain fixed, and identical joint actions result in identical travel times. We model non-determinism by varying SUMO's random seed, allowing identical route choices to produce different travel times for each agent, as in real-world traffic networks.

\subsection{Problem formulation}
\label{sec:problem_formulation}

We abstract the daily (repeated) route choices made by humans (and AVs in the future) in capacitated networks (limited resources) as a repeated congestion game \cite{holzman1997strong}. This can be formalized as a one-cycle Agent Environment Cycle (AEC) problem \cite{terry2021pettingzoogymmultiagentreinforcement}, defined in \cite{akman2025urburbanrouting, psarou2025collaborationcitymachinelearning} as a tuple:

\begin{equation*}
\left\langle 
   \mathcal{S}, \mathcal{I}, 
   \{\mathcal{U}_i\}_{i \in \mathcal{I}},
   \{r_i\}_{i \in \mathcal{I}},
   \{\Omega_i\}_{i \in \mathcal{I}},
   \upsilon
\right\rangle
\end{equation*}

We consider that in each episode (day or iteration) $\mathcal{I}$ AV agents in order $\upsilon$ of departure time, choose a route from their action space. We consider:



\begin{itemize}
\item A finite set of states $\mathcal{S}$.
\item  A finite set of agents $\mathcal{I}$ such that $N = |\mathcal{I}|$ ($\mathcal{H}$ is the set of human agents)
\item For each agent $i \in \mathcal{I}$:
\begin{itemize}
    \item The agent selects an action $u_i \in U_i$, forming a joint action $\mathbf{u} \in \mathbf{U} \;\equiv\; U^N$.
    \item $r_i$ is the immediate reward agent $i$ aims to maximize, as defined in \cite{akman2025urburbanrouting, psarou2025collaborationcitymachinelearning}.
    \item $\Omega_{i}$ is the set of possible observations for agent $i$.
    
\end{itemize}

\item $\upsilon$ is the agent selection mechanism. Agents act sequentially in the order of their departure times, so that those with earlier departure times act first.

\end{itemize}

In this problem, AVs learn their policies $\pi_i$, over episodes (days), to maximize their immediate rewards. For instance, like in examples below, they employ reinforcement learning to train policies (over consecutive days simulated in SUMO) to choose best actions (routes) in a given their current observation $\omega_i$:

\begin{equation}
\label{eq:policy}
   \pi_i(u_i \mid \omega_i)
\end{equation}

\subsection{Intrinsic reward signal}
\label{sec:intrinsic_reward_signal}

In the proposed method, we demonstrate that incorporating the marginal travel time of other agents into the intrinsic reward of each AV, in the context of the route choice problem, enables the system to converge faster towards user equilibrium. Following the work of \cite{jaques2019socialinfluenceintrinsicmotivation}, the reward of an agent $j$ is formulated as:
\begin{equation}
r_j(\mathbf{u}) = \alpha e_j(\mathbf{u}) + \beta m_j(\mathbf{u}), j \in \mathcal{I}
\label{eq:immediate_reward}
\end{equation}
where $e_j$ is the extrinsic environmental reward and $m_j$ is the influence reward of an agent $j$ and $\alpha$, $\beta$ are the weighting coefficients of each of these components.

\subsection{Marginal travel time}
\label{sec:mtt}

In this study, we followed the counterfactual definitions of the marginal travel time discussed in section \ref{sec:related_works} and formulated it as the change in the travel time of agent $i$ that arises when agent $j$ is removed from the system, compared to when agent $j$ remains present in the simulation. We define the marginal cost matrix as $M_{i, j}(\mathbf{u})$, a square matrix where rows represent an AV agent $i$ and columns represent an AV agent $j$. The entry at position $(i, j)$ denotes the travel time difference experienced by agent $i$ when agent $j$ is present compared to when agent $j$ is removed from the system. The travel time of agent $i$ when agent $j$ is removed from the system is indicated as $e_i(s, \mathbf{u}^{-j})$. The intrinsic reward for agent $j$, denoted as $m_j(\mathbf{u})$, is defined as the sum of values in column $j$ of the marginal cost matrix, representing the total impact agent $j$ has on the travel times of others. This impact is then transformed using the hyperbolic tangent function to produce a bounded, sign-preserving score that limits the influence of extreme values while preserving the direction of the effect.

\begin{equation}
m_j(\mathbf{u}) =
\sum_{\substack{i=1 \\ i \neq j}}^{N}
\tanh\!\left( e_i(\mathbf{u}) - e_i(\mathbf{u}^{-j}) \right)
\label{eq:intrinsic_reward}
\end{equation}

\begin{equation}
m_j(\mathbf{u})
= \sum_{i=1}^{N} M_{i,j}(\mathbf{u}), \quad N = |\mathcal{I}| \ \text{or}\ N = |\mathcal{H} \cup \mathcal{I}|
\label{eq:intrinsic_reward_part2}
\end{equation}


To construct the marginal cost matrix $M_{i, j}(\mathbf{u})$, we rerun (for each joint action $\mathbf{u}$) simulation $N$ times, once for each removed AV agent, while keeping the joint action and SUMO random seed fixed. In each run, a single agent $j$ is removed from the environment, and the resulting changes in travel times for all other agents are recorded. These values populate column $j$ of the marginal cost matrix, $M_{i,j}$, as in Algorithm \ref{alg:marginal_cost}. As a result, constructing the full matrix for one episode (one joint action) requires generating $N$ additional and separate environment instances for each joint action. For example, in the TRY network depicted in Fig. \ref{fig:overview}d, which has 2 discrete actions (routes), a scenario with 10 agents yields 1024 joint actions. If each joint action is encountered during training, an additional $1024\times 10$ environment runs will be needed.

To better understand the observed phenomenon, we consider two scenarios for calculating marginal cost. In the first scenario, AVs incorporate in their intrinsic reward the impact they have on the other AVs of the system, referred to as \textit{AV group marginal} (eq. \ref{eq:intrinsic_reward_part2}). To better understand how AVs can affect the total system, we incorporate the \textit{system marginal} (eq. \ref{eq:intrinsic_reward_part2}), which is the impact of an AV on all the drivers in the system, including both humans and AVs, denoted as \textit{system marginal}. 



Notably, when the intrinsic part of the reward (eq. \ref{eq:immediate_reward}) is computed as above, we obtain the interplay between the two often contradicting possible states in the urban mobility systems. The first part represents the natural, selfish inclination of utility-maximizing travelers to arrive at the destination as quickly as possible, while the latter represents the ideal cooperation to make the system as efficient as possible (see more in \cite{hoffmann2025wardropiancyclesmaketraffic}). Mixing the two in one formula may lead to interesting trade-off solutions. Here, we exploit this property in a system where user equilibrium (UE) is also system-optimal (SO) (which is rarely the case in complex systems), we sketch a proof suggesting that the resulting solutions will be stable somewhere between SO and UE in section \ref{sec:equilibria}.

\begin{algorithm}[!t]
\caption{Marginal cost matrix calculation for joint action $\mathbf{u}_T$}
\label{alg:marginal_cost}
\begin{algorithmic}[1]
\Require Simulation run $T$, parameters $\Theta$, agent set $\mathcal{I}$, joint action $\mathbf{u}_T$, travel times  $t_T$
\Ensure Marginal cost matrix $M$

\State $M_{i,j}(u) \gets \mathbf{0}^{\mathcal{N}\times \mathcal{N}}$

\For{$j \in \mathcal{I}$}  
    \State $\mathcal{I}^{(-j)} \gets \mathcal{I} \setminus \{j\}$ \Comment{Remove agent $j$}
    \State $\mathbf{u}_T^{(-j)} \gets \mathbf{u}_T \setminus \{u_j\}$
    \State $\mathrm{env} \gets \mathrm{TrafficEnvironment}(\mathcal{I}^{(-j)}, \Theta)$
    \State Run simulation $\mathcal{E}$ with $\mathrm{env}$ under actions $\mathbf{u}_T^{(-j)}$

    \For{$i \in \{1,\dots,A_m\}$}
        \State $t_{\mathcal{E}}(i) \gets$ travel time of agent $i$ in run $\mathcal{E}$
        \State $M_{i,j}(u) \gets 
            t_{\mathcal{E}}(i) - t_T(i)$
    \EndFor
\EndFor

\State \Return $M_{i, j}(u)$
\end{algorithmic}
\end{algorithm}

\section{Results}

We begin by introducing the minimal experimental setting employed in this work in section \ref{sec:experimental_setting}. Next, in section \ref{sec:try_net}, we present the main finding of this study: \textit{our proposed reward formulation achieves convergence to the optimal solution.} Afterwards, we examine the impact of different $\beta$ values on the convergence of the MARL algorithms in section \ref{sec:impact_beta}. In section \ref{sec:equilibria}, we show that under certain conditions the equilibria of the system remain unchanged despite increasing in the $\beta$ coefficient. Finally, in section \ref{sec:real_world_network}, We replicate our findings for the real-world Saint-Arnoult network, demonstrating that, after a short training, 50\% of the AV agents achieve shorter travel times, as well as shorter system travel times, when the marginal cost is introduced.

\begin{figure}[!h]
\begin{center}
\centerline{\includegraphics[width=\columnwidth]{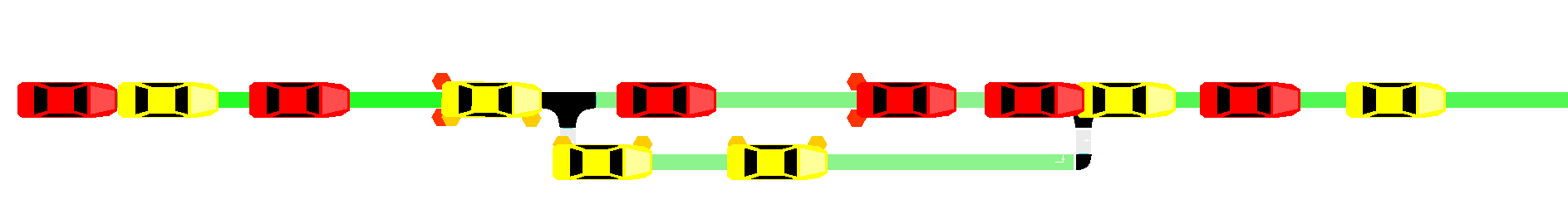}}
\caption{Snapshot from SUMO of the TRY network. Red vehicles represent human agents and yellow ones AVs.}
\label{fig:sumo_try_network}
\end{center}
\end{figure}

\begin{table}[!t]
\centering
\caption{Average travel times during the testing phase, last 100 episodes of the experiment, (in seconds) for each subgroup (AVs, humans), with standard deviations in parentheses.}
\label{tab:avg_travel_times}
\resizebox{\columnwidth}{!}{
\begin{tabular}{@{}l l l c c@{}}
\toprule
\multirow{2}{*}{Scenario} & \multirow{2}{*}{Reward type} & \multirow{2}{*}{Algorithms} & \multicolumn{2}{c}{Travel times}\\
\cmidrule(lr){4-5}
 &  &  & AVs & Humans\\
\midrule
\multirow{9}{*}{\rotatebox[origin=c]{90}{\textbf{Deterministic}}}
 & \multirow{3}{*}{Selfish AVs}     & UCB  & 65.7 (15.6) & 65.1 (24.3) \\
 &                                  & MAPPO & 62.0 (14.8) & 64.4 (27.3)\\
 &                                  & IDQN & 64.1 (14.5) & 70.3 (28.6) \\
\cmidrule(lr){2-5}
 & \multirow{3}{*}{AV group marginal} & UCB  & 57.9 (12.0) & 51.7 (14.4) \\
 &                                  & MAPPO & 57.8 (12.4) & 51.9 (14.8)\\
 &                                  & IDQN & \textbf{56.8 (11.5)} & \textbf{50.9 (13.6)} \\
\cmidrule(lr){2-5}
 & \multirow{3}{*}{System marginal}   & UCB  & 57.9 (12.6) & 51.2 (14.5) \\
 &                                  & MAPPO & \textbf{57.7 (12.2)} & \textbf{51.1 (14.3)}\\
 &                                  & IDQN & 59.2 (12.0) & 51.7 (13.4) \\
\midrule
\multirow{9}{*}{\rotatebox[origin=c]{90}{\textbf{Non Deterministic}}}
 & \multirow{3}{*}{Selfish AVs}     & UCB  & 62.9 (14.4) & 61.5 (19.8) \\
 &                                  & MAPPO & 62.0 (15.1) & 64.3 (27.5)\\
 &                                  & IDQN & 63.7 (14.6) & 67.7 (27.8) \\
\cmidrule(lr){2-5}
 & \multirow{3}{*}{AV group marginal} & UCB  & \textbf{56.9 (11.7)} & \textbf{51.0 (13.8)} \\
 &                                  & MAPPO & 57.0 (11.7) & 51.0 (13.9)\\
 &                                  & IDQN & 59.1 (12.4) & 52.1 (13.8) \\
\cmidrule(lr){2-5}
 & \multirow{3}{*}{System marginal}   & UCB  & \textbf{57.0 (12.0)} & \textbf{50.9 (13.8)} \\
 &                                  & MAPPO & 57.0 (11.7) & 51.0 (13.9)\\
 &                                  & IDQN & 59.7 (12.1) & 52.0 (13.9) \\
\bottomrule
\end{tabular}
}
\end{table}

\begin{figure*}
    \centering
    \includegraphics[width=1\linewidth]{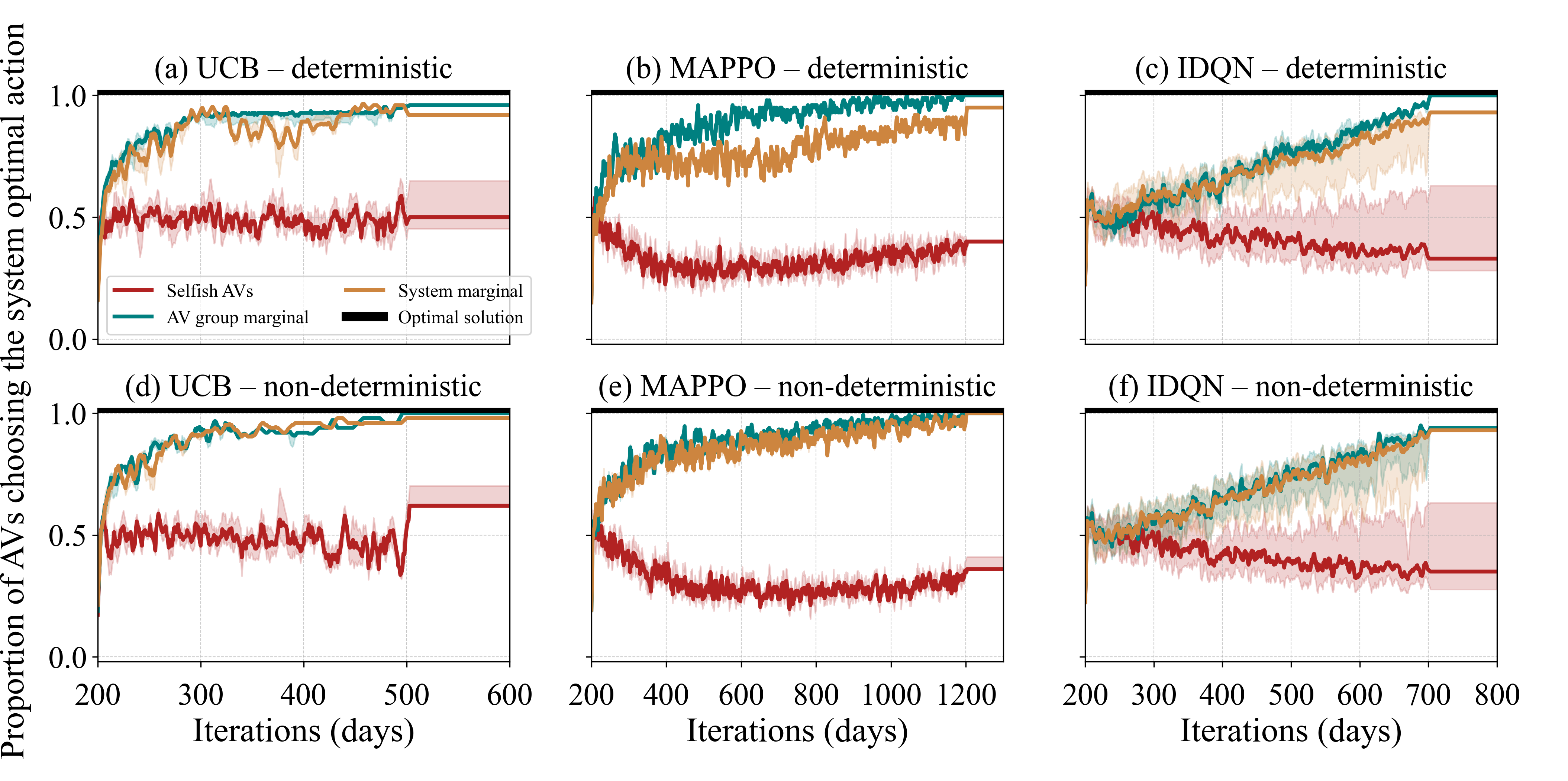}
    \caption{Introducing the marginal travel time in the reward of the AV agents accelerates convergence of the AVs to the optimal solution. We demonstrate this by incorporating two types of marginal travel times (MTTs): one that considers the impact on AV agents, \textit{AV group marginal}, and the second that considers the impact on all drivers, \textit{system marginal} (eq. \ref{eq:intrinsic_reward_part2}). We also demonstrate that our method enhances convergence in non-deterministic traffic dynamics, which more accurately represent real-world traffic conditions. 
    When the proportion of AVs choosing the optimal action is close to 1, it indicates that nearly all agents selected the optimal solution. A proportion of 0.5 indicates that half of the agents chose the optimal option. The last 100 iterations of the plots depict the evaluation mode, where agents use the learned policy without exploration.}
    \label{fig:try_network_results}
\end{figure*}

\subsection{Experimental setting}
\label{sec:experimental_setting}

To demonstrate the impact of the intristic reward on the solution and training, we select the carefully tailored minimal toy network example, used in all the experiments of this work, except for section \ref{sec:real_world_network} where we experiment with more complex and realistic setting. In the TRY network (illustrated in Fig. \ref{fig:sumo_try_network}), agents everyday have the option to choose between two routes: Route 0, which is faster but does not have priority (yield sign), and Route 1, which is slightly longer but has priority. These experiments involve 22 agents. 
Initially, all are human drivers who learn their route-choice behavior for 200 days (iterations). During this learning phase, they converge to consistently selecting Route 0 (which, by design, is both Wardrop equilibrium and System Optimum). 

Following the initial 200 episodes (days), 10 of the human drivers are assumed to switch to using AVs for their daily commute. We consider that AVs start learning their routing strategies from episode 200 using different MARL algorithms. We select most popular, promising and fitting state-of-the-art RL algorithms, building on findings from \texttt{URB} benchmark \cite{akman2025urburbanrouting} we use here: MAPPO \cite{mappo}, IDQN \cite{mnih2013playingatarideepreinforcement}, and UCB \cite{auer2002finite}. After episode 200, we assume human drivers' strategies remain fixed, reflecting the realistic assumption that human drivers typically settle on a perceived optimal route after repeated experience \cite{WARDROP1952}. For simplicity, we assume that remaining human drivers will not adapt to the changes induced by AVs and their training.

\subsection{Performance of the marginal cost reward formulation}
\label{sec:try_net}

In this section, we demonstrate that when running different MARL algorithms with our new reward formulation under both deterministic and non-deterministic traffic dynamics, the AV agents converge faster to the optimal solution, as depicted in Fig. \ref{fig:try_network_results}. Specifically, in Fig. \ref{fig:try_network_results}(a, b, d, e), after as little as 100 iterations, the agents reach a joint action that is close to the optimal one.

Additionally, we consider two scenarios of marginal cost calculation discussed in Section \ref{sec:mtt} into our demonstration in Fig. \ref{fig:try_network_results}. One where AVs account for their impact on the other AVs in the system, referred to as \textit{AV group marginal}, and the another where the intrinsic reward reflects the impact on all drivers in the system, denoted as \textit{system marginal}. As shown in Fig. \ref{fig:try_network_results}. We did not observe significant differences between the two formulations, both the \textit{AV group marginal} and the \textit{system marginal} require a similar amount of time to converge to the optimal solution.


\begin{table}[ht]
    \centering
    \caption{Sample from the marginal cost matrix for the 
    \textit{AV group marginal} case, computed at joint action \mbox{[1, 0, 1, 0, 0, 0, 1, 0, 1, 1]}, under deterministic traffic dynamics, for first three Avs (values in seconds). }
    \begin{tabular}{l|SSS}
        \hline
        ID & AV 1 & AV 3 & AV 5\\
        \hline
        AV 1 & 0 & 0 & 0 \\
        AV 3 & -21 & 0 & -9.6 \\
        AV 5 & 0 & -1.2 & 0 \\
        \hline
    \end{tabular}
    \label{tab:marginal_cost_matrix}
\end{table}

Table \ref{tab:avg_travel_times} shows the average travel times of AVs and human agents during the testing phase, where AVs utilize the policy they learned without including any exploration. In both deterministic and non-deterministic traffic settings, the average travel times of AVs and human drivers are shorter in both the \textit{AV group marginal} and the \textit{System marginal} across all the tested algorithms. By introducing marginal cost, travel times are reduced from over 70s (IDQN) to 57 seconds for AVs and to 51 for human drivers. Notably, AV agents achieve similar travel times in both marginal formulations (\textit{AV group marginal}, \textit{System marginal}), which are shorter compared to when the marginal cost is not introduced. These results support our claim that, when the marginal cost is introduced into the reward of AV agents, they converge faster to the system and individually optimal solutions. 

Table \ref{tab:marginal_cost_matrix}, samples values of the marginal cost matrix. Notably, the matrix is often not lower triangular, which indicates that the system does not satisfy the First-In-First-Out (FIFO) property. This occurs because the priority scheme of the yield sign allows later-departing agents to influence earlier-departing ones, thereby amplifying non-stationarity. For example, in Table \ref{tab:marginal_cost_matrix}, AV agent 3 selects route 0, which does not have priority, and AV agents 1 and 5 choose route 1. Agent 1 departs at timestep 1, agent 3 at timestep 3 and agent 5 at timestep 5. Consequently, agent 3 must yield to agents 1 and 5 in the intersection. Therefore, removing these two agents from the system would benefit agent 3 - which is reflected in the values of matrix $M$.

\subsection{Impact of the social component weight $\beta$}
\label{sec:impact_beta}

\begin{figure}[!t]
\begin{center}
\centerline{\includegraphics[width=\columnwidth]{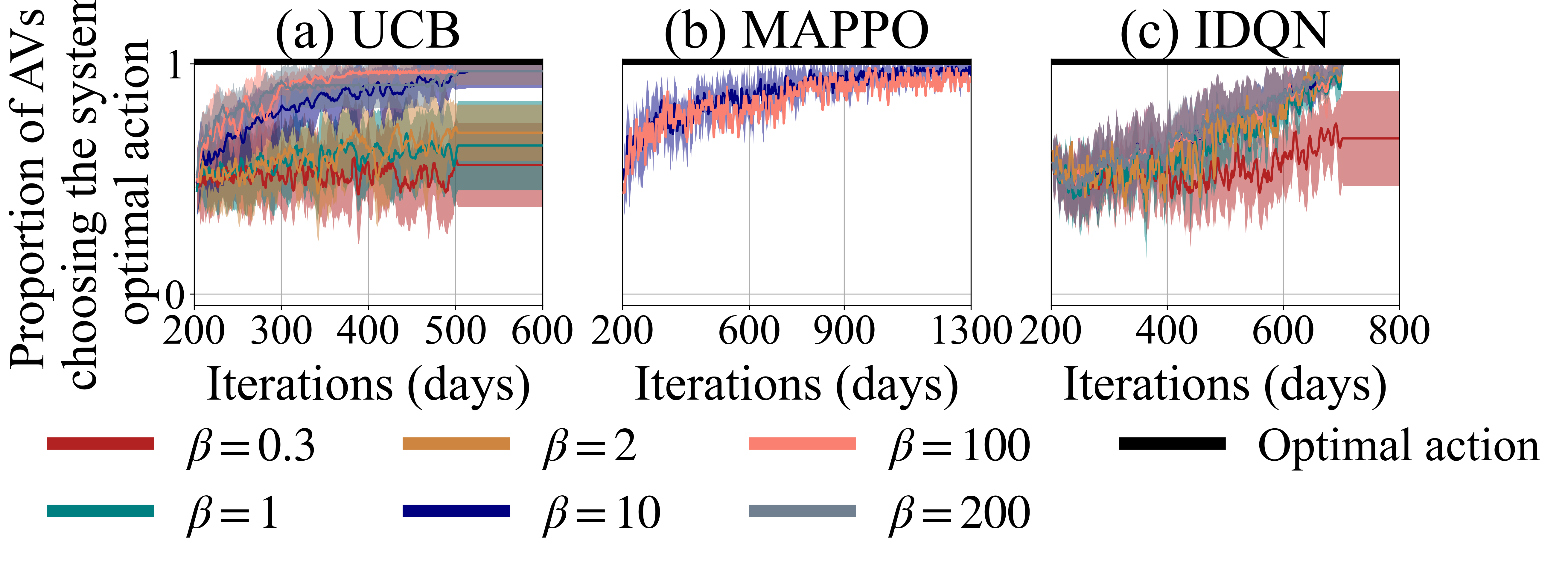}}
\caption{Effect of different values of the shaping coefficient $\beta$ on the convergence of the UCB (a), MAPPO (b), and IDQN (c) algorithms. Higher values of $\beta$ lead to faster convergence to the optimal solution. When the proportion of agents choosing the system optimal action is close to 1, it indicates that nearly all agents selected the optimal solution.}
\label{fig:reward_shaping}
\end{center}
\end{figure}

Here, we investigate how incorporating agents' marginal impact into the reward value influences the convergence of the learning algorithms. Fig. \ref{fig:reward_shaping} shows the convergence of the UCB \cite{auer2002finite}, MAPPO \cite{mappo}, and IDQN \cite{mnih2013playingatarideepreinforcement} algorithms when the coefficient of the intrinsic reward, $\beta$ takes different values. For UCB, $\beta$ equal to 10 results in convergence to the optimal solution after approximately 300 training episodes. Suggesting that higher values of $\beta$ (100 and 200) enable the agents to reach the optimal solution faster. 

For the MAPPO algorithm, however, increasing $\beta$ from 10 to 100 does not noticeably improve the convergence rate. For IDQN, all the experiments with the tested beta values converge to the optimal solution after approximately 400 iterations (expect from the ones with $\beta=0.3$). However, IDQN requires more iterations compared to UCB and MAPPO, where the system- and individually optimal solutions are nearly reached after roughly one-third of the total training, as shown in Fig. \ref{fig:reward_shaping}. Based on these findings, and in line with the experimental results presented in Fig. \ref{fig:reward_shaping}, we select $\beta=200$ for all the experiments shown in this paper, except for this section and section \ref{sec:real_world_network}. The varying number of selected training iterations for each algorithm indicates that, depending on the selected number of training iterations, tuning the $\beta$ value can facilitate convergence to the optimal solution. 


\subsection{Equilibria}
\label{sec:equilibria}



Now we analyze the equilibria that arise in the system when the intrinsic rewards described in eq. \ref{eq:intrinsic_reward_part2} are introduced into the rewards of the AV agents. Equilibrium solutions, and specifically the Nash equilibria, have been adopted as the standard solution concept in MARL \cite{marl-book}, we aim to understand how these equilibria can be affected by changes in the reward structure. Since any change in the equilibria can alter the agents’ optimal policies, and consequently their observed behaviors \cite{learning_to_drive_a_bicycle}.
Specifically, we investigate how varying the coefficients of the intrinsic and extrinsic rewards ($\alpha, \beta$), as defined in eq. \ref{eq:immediate_reward}, influences the resulting equilibria. 

To formally determine whether a joint action is a Nash equilibrium, we test the $2^{10}$ (1024) possible joint actions. For each of these possible joint actions $\textbf{u} \in U^{10}$, we test unilateral deviations. A joint action is a Nash equilibrium if no agent can increase its reward, sum of the negative travel time of the agent and the MTT, by changing its own action while the decisions of all other agents remain fixed. When assessing a unilateral deviation in the Nash equilibrium calculation, an agent $j$ compares whether it is better for him to choose the alternative action (route 1 if the initial action was route 0 or route 0 if the initial action is route 1 in the TRY network).

We first analyze the case where agents receive no intrinsic reward and aim to minimize their own travel times $(\alpha = 1, \beta = 0)$. In this setting, the system has a unique equilibrium in which all agents choose route 0. This corresponds to a \textit{Wardrop Equilibrium}  \cite{Wardrop1952ROADPS}, the analogue of the Nash equilibrium in the route choice context, and in our specific setting, a \textit{System Optimum}, as the total travel time across the network is minimized in that state \cite{merchant1978optimality}.

\begin{figure}[!t]
\begin{center}
\centerline{\includegraphics[width=\columnwidth]{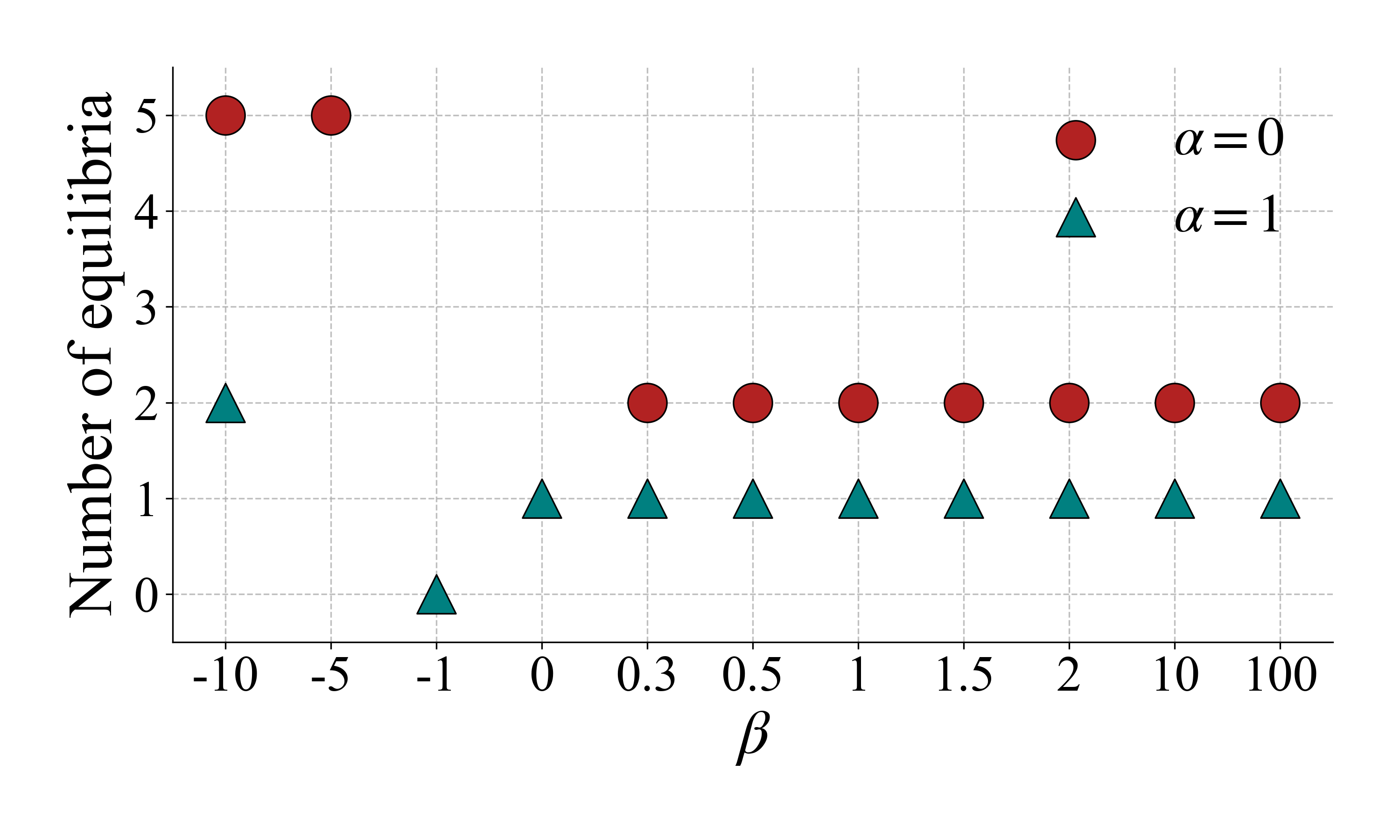}}
\caption{Number of equilibria in the system for different $\alpha$, $\beta$ coefficients. When $\beta > 0$ and $\alpha = 1$, the system has one unique equilibrium, the same as the equilibrium when the AV agents are selfish ($\beta = 0$ and $\alpha = 1$). This suggests that with $\beta$ values above a certain threshold and $\alpha = 1$, the equilibrium state remains unchanged, as discussed in section \ref{sec:equilibria}. Additionally, for hypothetical values of $\beta < 0$ (malicious behavior, \cite{jamroz2025social}),  multiple equilibria become possible in the system. }
\label{fig:equilibria}
\end{center}
\end{figure}

We next explore how the introduction of intrinsic rewards and the variation of $\beta$ affect the equilibrium structure (Fig. \ref{fig:equilibria}). Starting with $\alpha = 1$, $\beta = 0.3$, we observe that the equilibrium remains unchanged, compared to the state where $\alpha = 1$, $\beta = 0$. \textbf{Increasing $\beta$ up to 100 leads to the same outcome: the equilibrium state is unaffected and it remains unique. }

\subsubsection*{Proof}

One of our claims is that introducing the counterfactual marginal cost-based intrinsic reward (eq. \ref{eq:intrinsic_reward}) improves convergence without altering the underlying game. If the intrinsic reward were to shift the equilibrium, then our improvement would be dismissed as optimizing a different problem. Therefore, it is crucial to show that increasing the coefficient $\beta$ beyond a certain threshold does not affect the equilibrium state. 
The proof described in this section concerns the TRY network, depicted in Fig. \ref{fig:sumo_try_network} but, can easily be extended to bigger networks.

The counterfactual intrinsic reward for an agent $j$, as described in eq. \ref{eq:intrinsic_reward} is obtained by re-simulating the system, once with the agent $j$ present and once where agent $j$ is removed. Since agent $j$ is removed from the simulation in this counterfactual no matter which route it would have chosen, the travel times of the remaining agents are the same whether $j$ chose route 0 or route 1: 
\begin{equation*}
\sum_{\substack{i=1 \\ i \neq j}}^{N} e_i\!\bigl( \mathbf{u}^{-j}|_{u_j = 0}\bigr)
\;=\;
\sum_{\substack{i=1 \\ i \neq j}}^{N} e_i\!\bigl( \mathbf{u}^{-j}|_{u_j = 1}\bigr), \quad N = |\mathcal{I}| \ 
\end{equation*}

\textit{Unilateral deviation \(u_j : \mathbf 0 \to \mathbf 1\) (with \(\mathbf u^{-j}\) fixed).}
Let's consider the deviation of agent $j$ from its chosen action while keeping the actions of all the other agents fixed. The only component of the intrinsic reward that is affected by agent $j$ deviating is the travel times of all the individual agents when agent $j$ chooses action 0 or 1. Hence, adding the marginal-cost term contributes an action-invariant constant and rescales the same $
\sum_{\substack{i=1,\, j \neq i}}^{N} \left( e_i\!\left(\mathbf{u}^{-j}\right) \right)$ by a scale $\beta$, leaving the ordering of routes, and therefore best responses and the Nash equilibrium unchanged as $\beta$ increases. 

Agent \(j\) minimizes their reward: $r_j( \mathbf u)\;=\;e_j( \mathbf u)\;+\;\beta\,m_j( \mathbf u)$. To verify whether the equilibrium is preserved, it is sufficient to examine the effect of a single agent unilaterally switching its route while all the other agents keep their choices fixed. We then define the own travel-time change of agent $j$ and the corresponding system-wide change as follows:

\begin{align*}
\delta_j \triangleq e_j\!\big(\mathbf u^{-j}\big|_{u_j=1}\big) - e_j\!\big(\mathbf u^{-j}\big|_{u_j=0}\big), \qquad
\Delta C_j \triangleq \sum_{\substack{i=1 \\ i \neq j}}^{N} \tanh(\delta_i).
\end{align*}
Agent's \(j\in \mathcal{I}\) reward change when switching from \(\mathbf 0\) to \(\mathbf 1\) is

\begin{equation*}
\Delta r_j(\beta)
\;\triangleq\;
r_j\!\big(\mathbf u^{-j}\big|_{u_j=1}\big)
\;-\;
r_j\!\big(\mathbf u^{-j}\big|_{u_j=0}\big)
=\!\!\!\!\!
\underbrace{\delta_j}_{\text{own change}}
\!\!\!\!\!
+\;
\beta\!\!\!
\underbrace{\Delta C_j}_{\text{system change}}
\!\!\!\!\!\!.
\end{equation*}

\textit{Invariance of best response (route 0 better for all \(\beta\ge 0\)).}

Because \(\tanh\) is odd and strictly increasing, we have \(\operatorname{sign}(\tanh(\delta_i))=\operatorname{sign}(\delta_i)\).
If in addition, we assume aligned externality, i.e., if \(\operatorname{sign}(\Delta C_j)=\operatorname{sign}(\delta_j)\) then for every \(\beta\ge 0\) holds \(\operatorname{sign}(\Delta r_j(\beta))=\operatorname{sign}(\delta_j)\) . Indeed, if \(\delta_j > 0\) and \(\Delta C_j \geq 0\), then for every \(\beta \geq 0\) we have \(\Delta r_j(\beta) = \delta_j + \beta \Delta C_j \geq \delta_j > 0\), for \(\delta_j < 0\) and \(\Delta C_j \leq 0\), we get for every \(\beta \geq 0\) that \(\Delta r_j(\beta) = \delta_j + \beta \Delta C_j \leq \delta_j < 0\) and if \(\delta_j = 0 = \Delta C_j\), hence \(\Delta r_j(\beta) = 0\) for all \(\beta\).
More generally, on a finite range \(\beta\in[0,\beta_{\max}]\), where \(\beta_{\max}= \sup\{\beta\geq0:\Delta r_j(\beta) > 0\}\)
\begin{equation}\label{eq:best_response_invariance}
\Delta r_j(\beta) > 0 \ \ \forall\,\beta\in[0,\beta_{\max}]
\quad\Longleftrightarrow\quad
\Delta C_j \;>\; -\,\frac{\delta_j}{\beta_{\max}}.
\end{equation}
Fix \(\beta_{\max} > 0\). Because \(r_j(\beta)\) is affine in \(\beta \in [0, \beta_{\max}]\), its minimum occurs at an endpoint. If \(\Delta C_j \geq 0\), \(\Delta r_j\) is nondecreasing so its minimum is at \(\beta = 0\). Thus \[\Delta r_j(\beta)>0 \ \forall \beta \in [0, \beta_{\max}] \Longleftrightarrow \Delta r_j(0) = \delta_j > 0.\]
If \(\Delta C_j < 0\), \(\Delta r_j\) is strictly decreasing so its minimum is at \(\beta = \beta_{\max}\). Thus \[\Delta r_j(\beta) > 0 \ \forall \beta \in [0, \beta_{\max}] \Longleftrightarrow \Delta r_j (\beta_{\max}) = \delta_j + \beta_{\max} \Delta C_j > 0.\] 

Therefore, when \(\beta \in [0, \beta_{\max}]\), the equilibria of the system remain intact. 

\subsection{Real-world network}
\label{sec:real_world_network}

To test our claim that incorporating the marginal cost into the reward of the AV agents enhances system-wide performance and can reduce the travel times of individual AV agents, we present results from a bigger real-world network where the SO and UE solutions do not coincide. Specifically, we apply our new reward formulation to the real-world network of Saint-Arnoult from the URB benchmark \cite{akman2025urburbanrouting}, which includes 1289 nodes and 2011 edges. In the experiment, we consider 111 traveling agents, with each agent selecting from an action space of three available routes (leading to a huge joint action space $3^{110}$). We consider that after period of human training, 10 of the human agents switch to traveling with AVs, learning the optimal route choice for 300 iterations using the UCB algorithm, and subsequently evaluating the learned policy for an additional 10 iterations. Notably, now (unlike in TRY network) the UE and SO solutions do not coincide.

\begin{table}[h!]
\caption{Travel times (in seconds) of 10 AVs in Saint-Arnoult with selfish objective and our proposed marginal intristic reward. 
Even in this short training experiment, for 6 out of 10 AVs using marginal cost in their reward formulation ($\beta = 0.3$) allowed to reduce their individual travel times (selfish rewards). Both the system-wide and AV group travel times were improved.}
\label{tab:av_travel_times_saint_arnoult}
\centering
\resizebox{\columnwidth}{!}{%
\begin{tabular}{cccc}
\hline
AV ID & Selfish AVs & AV group marginal & System marginal\\
\hline
2   & 125.17 (0.44)  & 126.95 (0.11) &  127.01 (0.18)\\ 
37  & 116.56 (1.14)  & 117.59 (0.84) &  117.55 (0.04) \\
80  & 112.43 (0.36) & 113.24 (1.55)  &  \textbf{112.40 (0.16)}\\
52  & 132.54 (1.93)  & \textbf{132.36 (0.02)} &  \textbf{132.36 (0.7)}\\
107 & 165.10 (0.33)  & 163.04 (0.13)          &  \textbf{163.41 (0.02)}\\
4   & 181.86 (0.53)  & 183.57 (0.14)          &  \textbf{180.12 (0.44)}\\
95  & 192.98 (0.74) & 193.06 (0.23) &  193.86 (1.27)\\
74  & 215.28 (0.82)  & \textbf{213.25 (0.71)} &  \textbf{212.51 (0.22)}\\
102 & 213.56 (0.35)  & 214.88 (0.24)         &  215.20 (0.57)\\
34  & 400.33 (1.08)  & \textbf{393.61 (0.05)} &  \textbf{393.24 (0.13)}\\
\hline
System   & 27493.92 (15.55) & \textbf{27494.81 (2.96)} &  \textbf{27488.84 (0.74)}\\
AV group & 1855.05 (2.55)  & \textbf{1851.31 (1.58)} &  \textbf{1847.85 (3.37)}\\
\hline
\end{tabular}%
}
\arrayrulecolor{black} 
\end{table}

In Table \ref{tab:av_travel_times_saint_arnoult}, we compare the average travel times (among replications) of each individual AV agent, of the overall system, and of the AV group when the training is complete. The comparison is made under three reward schemes for AVs: the selfish travel-time–based reward, and the selfish travel-time–based reward including the \textit{AV group marginal}, and the \textit{System marginal} (eq. \ref{eq:immediate_reward}, \ref{eq:intrinsic_reward_part2}). We observe that with a relatively short training (300 iterations, i.e. as much as in the simple TRY network, whereas we are now in a much more complex setting), more than 50\% of the agents achieve shorter individual travel times when the marginal cost is introduced, as shown in Table \ref{tab:av_travel_times_saint_arnoult}. Additionally, the system and the AV group average travel times are reduced when we introduce the social component in the reward of the AVs. This result emphasizes that incorporating socially oriented behavior in AVs can be efficient in improving both individual and system-level performance of future transportation systems.

\section{Discussion and Conclusions}

In this work, we demonstrated that introducing socially aware MARL-based AV agents into future traffic networks can help relief the convergence challenges that arise when AVs act selfishly. Specifically, we complement the selfish travel-time-based reward with a counterfactual, equilibrium-preserving intrinsic reward based on the marginal cost that each AV imposes on the other drivers of the system. We demonstrate that our novel reward formulation not only improves the overall system performance but, surprisingly, is beneficial for the individual AV agents. Additionally, we show that the proposed reward formulation preserves the system's equilibria while accelerating convergence toward the system-optimal solution. 

In our experimental evaluation, we compare AV agents trained using MARL algorithms and optimizing different objectives: our proposed reward formulation versus the selfish travel-time based reward. On the simple TRY network, we demonstrate that socially aware AVs converge faster to the optimal solution, which in this case coincides with the individually optimal solution. In the real-world Saint-Arnoult network, after a short training period, more than 50\% of the socially aware AVs achieve shorter individual travel times, and the overall system travel times are also improved. 

A limitation of this work includes the computational cost of calculating the marginal cost matrices, as they will grow exponentially with size of the joint action space. Moreover, we deliberately focused on the rare case where SO and UE coincide. Future studies shall better exploit how including the social component may be both effective in reducing Price of Anarchy and remaining attractive and fair for AV users.

This work can contribute towards demonstrating the importance of introducing some form of socially-aware collaboration between AV agents in future traffic systems. Understanding the potential impact of introducing AVs into future cities is of paramount importance, particularly when it comes to identifying potential negative effects and examining whether some form of collaboration between the vehicles could help maintain overall system stability, equity and performance.






\newpage
\bibliographystyle{ACM-Reference-Format} 
\bibliography{sample.bib}


\newpage

\end{document}